\documentclass[%
 preprint,
amsmath,amssymb,
aps,
]{revtex4-1}

\usepackage{graphicx}
\usepackage{dcolumn}
\usepackage{bm}

\newcommand{\Tr}{\mathop{\rm Tr}\nolimits}

\begin{document}


\title{The elastic Maier-Saupe-Zwanzig model and some properties of nematic elastomers}

\author{Danilo B. Liarte}
\email[]{danilo@if.usp.br}
\author{Silvio R. Salinas}
\email[]{ssalinas@if.usp.br}
\author{Carlos S. O. Yokoi}
\email[]{cyokoi@if.usp.br}
\affiliation{Instituto de F\'isica, Universidade de S\~ao Paulo\\Caixa Postal 66318, CEP 05314-970 S\~ao Paulo, SP, Brazil}

\date{\today}

\begin{abstract}
  We introduce a simple mean-field lattice model to describe the
  behavior of nematic elastomers. This model combines the
  Maier-Saupe-Zwanzig approach to liquid crystals and an extension to
  lattice systems of the Warner-Terentjev theory of elasticity, with
  the addition of quenched random fields. We use standard
  techniques of statistical mechanics to obtain analytic solutions for
  the full range of parameters. Among other results, we show the
  existence of a stress-strain coexistence curve below a freezing
  temperature, analogous to the $P$-$V$ diagram of a simple fluid, with
  the disorder strength playing the role of temperature. Below a
  critical value of disorder, the tie lines in this diagram resemble
  the experimental stress-strain plateau, and may be interpreted as
  signatures of the characteristic polydomain-monodomain transition.
  Also, in the monodomain case, we show that random-fields may soften
  the first-order transition between nematic and isotropic phases,
  provided the samples are formed in the nematic state.
\end{abstract}

\pacs{Valid PACS appear here}
\maketitle


\section{Introduction}

Liquid-crystalline molecules imbedded in a polymer network give rise
to the novel class of elastomer systems, with coupled rubber
elasticity and orientational order, and rather unusual properties
\cite{warner03,warner96}. Nematic elastomers (NEs) may undergo a
distortion in response to an alignment of the nematic units as the
sample is cooled below the nematic-isotropic transition temperature
$T_{\text{NI}}$.  Reciprocally, the application of an external stress
may give rise to nematic ordering of an initially disordered sample.
There is a number of suggestions of applications of these new
soft-matter materials, ranging from uses in optics (in bifocal lenses,
for instance) to applications as thermo-mechanical devices
\cite{warner03}.

Compared with conventional liquid crystals, NEs present a peculiar
transition from the nematic to the isotropic states.  In usual
nematics, according to the Landau-de Gennes theory, symmetry
requirements lead to a first-order transition, with a jump of the
nematic order parameter, at a transition temperature $T_{\text{NI}}$.
A different scenario is observed in the NEs.  Instead of a
discontinuity there is a continuous but quick variation of the order
parameter during the transition. The microscopic mechanism behind this
non-trivial behavior has been discussed by many authors
\cite{uchida00,selinger02,selinger04,petridis06,jayasri09}.  In
particular, continuous three-dimensional coarse-grained theories
\cite{petridis06}, and numerical simulations of microscopic models
\cite{selinger04}, indicate that quenched random-field interactions
may smooth out the characteristic first-order transition in these
systems. In NEs these random fields are supposed to originate from internal stresses
produced by the network cross-links \cite{fridrikh97,yu98}. More recently, slightly different random-field interactions have been suggested by Lu et al. to describe network heterogeneity in randomly crosslinked materials \cite{lu11}.

Random fields may also be relevant to describe the interesting
transition from states of polydomain to a monodomain in NEs
\cite{fridrikh99}. If the system is cooled below $T_{\text{NI}}$, NEs
are known to display a stable structure, the so-called
\textquotedblleft Schlieren texture\textquotedblright, characterized
by sets of frozen randomly-oriented domains of mesogen units. This
polydomain state may be turned into a monodomain state by stretching
the sample under uniaxial tension. For a certain range of low
temperatures, experiments show an unusual stress-strain curve with
three characteristic regions. In the first region, for small strain,
the system is in the polydomain state, the sample is opaque, and
stress increases linearly with strain, according to Hooke's law. In
the second region, the stress is constant for a range of intermediate
values of strain, and there is a dramatic increase of the nematic
order parameter. This plateau is then followed by an increase of
stress at larger strains, with the system in the optically transparent
monodomain state \cite{clarke98,ortiz98,ortiz98b}. Some authors have
suggested that cross-linking conditions are essential to explain the
stable polydomain state \cite{fridrikh99,uchida00}. In particular,
random-fields and the neoclassical theory of elasticity have been used
by Fridrikh and Terentjev to obtain a good fitting of some
experimental findings \cite{fridrikh99}.

In this article we propose a lattice statistical model for nematic
elastomers, at the mean-field level, which can be analytically solved
in the presence of stress and random fields. This model has the
advantage of providing a simple way to investigate a wide range of
parameters, gives a unified view of the critical behavior of NEs, in
general agreement with previous theoretical
\cite{fridrikh99,petridis06}, numerical \cite{selinger04}, and
experimental findings \cite{warner03}, and suggests the occurrence of
some additional phenomena. In Section II, we define the basic lattice
model. The global free energy is obtained in Section III. Some
specific calculations, including stress-strain curves, comparisons
with the literature, and a few new predictions are presented in
Section IV. The main conclusions are given in the last Section.

\section{Lattice model for a nematic elastomer}

The Maier-Saupe (MS) model, which is known to provide a good
description of the isotropic-nematic transition
\cite{maier58,gennes93}, is the liquid-crystalline analogue of the
Curie-Weiss model of ferromagnetism \cite{salinas93}. The idea
consists in the extension of the finite-range interactions of
realistic systems to infinite-range interactions in order to construct
a simpler model, which turns out to be exactly soluble. In the MS
model, the basic elements are the molecular unit vectors $\bm{n}_{i}$,
representing a set of $N$ mesogens, which interact via the quadrupole
energy
\begin{equation}
  E_{\text{MS}}=-\frac{A}{N}\sum_{1\leq i<j\leq N}\sum_{\mu,\nu=x,y,z}
  S_{i}^{\mu\nu}S_{j}^{\mu\nu},
  \label{emseq1}
\end{equation}
where $A>0$ is an energy constant,
\begin{equation}
  S_{i}^{\mu\nu}=\frac{1}{2}(3n_{i}^{\mu}n_{i}^{\nu}-\delta^{\mu\nu}),
\end{equation}
and $|\bm{n}_{i}|=1$ for $i=1,2,\cdots,N.$ The calculations may be
further simplified if we adopt a suggestion of Zwanzig and restrict
the number of allowed orientations of $\bm{n}_{i}$ to the six values
along the Cartesian axes,
\begin{equation}
  \bm{n}_i \in \left\{ (\pm1,0,0),(0,\pm1,0),(0,0,\pm1)\right\}.
  \label{nzwanz}
\end{equation}
Generalizations of this model, which we call the Maier-Saupe-Zwanzig
(MSZ) model, have been applied to the study of biaxial and uniaxial
ordering in rod-plate mixtures of liquid crystals
\cite{henriques97,carmo10}.  In agreement with predictions of
Landau-de Gennes theory \cite{gennes93}, the isotropic-nematic
transition is found to be first-order, with a discontinuity in the
nematic order parameter. In the nematic elastomer case, Xing \emph{et
  al.} have considered similar Maier-Saupe type of interactions
adjunct to a microscopic model to make contact with a macroscopic
Landau theory \cite{xing08}.

To model the mechanical and orientational character of nematic elastomers, a number of approaches have been considered \cite{golubovic89,uchida00,warner03,selinger04,pasini05,ennis06,jayasri09}. We shall assume that elastic properties arise from an entropic
contribution \cite{warner03}. Thus we consider the canonical partition
function,%
\begin{equation}
  Z=\sum_{\{\bm{n}_{i}\}}\,\Omega\left( \{\bm{n}_{i} \},\Lambda\right)
  \exp\left( -\beta E_{\text{MS}}\right), 
\end{equation}
where $\beta=1/k_BT$, $E_{\text{MS}}$ is the interaction energy of the
MSZ model given by Eq. \eqref{emseq1}, and the sum is over the
configurations \eqref{nzwanz} of the microscopic nematic directors
$\bm{n}_{i}$. The entropic term $\Omega$ depends on the nematic
orientations and on a global lattice distortion tensor
$\bm{\Lambda}$ \cite{warner03}. If we consider a uniform strain
along the direction of a unit vector $\bm{m}$, the distortion
components may be written as
\begin{equation}
  \Lambda_{\alpha\beta}=\lambda^{-1/2} \delta_{\alpha \beta}+(\lambda-\lambda^{-1/2})m_{\alpha
  }m_{\beta},\quad\alpha,\beta=x,y,z.
\end{equation}
According to an extension of the neoclassical theory of elasticity
\cite{warner96,warner03} for lattice Hamiltonian systems, proposed by
Selinger and Ratna \cite{selinger04}, we write the degeneracy as
\begin{equation}
  \Omega=\exp\left( -\beta F_{\text{el}}\right),
\end{equation}
with the elastic free energy
\begin{equation}
  F_{\text{el}}=\frac{\mu}{2}\sum_{i}\Tr \left(\bm{l}_{0,i} \cdot
    \bm{\Lambda}^{T}\cdot\bm{l}_{i}^{-1}\cdot\bm{\Lambda}\right), 
  \label{elasticfe}
\end{equation}
where $\mu$ is the linear shear modulus, $\bm{l}_{i}$ is a local shape
tensor, and $\bm{l}_{0,i}$ is the local shape tensor at the time of
the cross-linking. The components of the shape tensors are obtained
from the equation
\begin{equation}
  l_{i,\alpha\beta}^{-1}=l_{\perp}^{-1}\delta_{\alpha\beta}+\left(
    l_{\parallel}^{-1}-l_{\perp}^{-1}\right) n_{i,\alpha}n_{i,\beta},
\end{equation}
where $l_{\perp}$ and $l_{\parallel}$ are the effective step lengths
of the nematic polymers in the perpendicular and parallel directions
with respect to the nematic vectors. If the cross-linked network is
formed with the sample in a totally disordered isotropic state, we
assume that the shape tensor $\bm{l}_{0,i}^{-1}$ is given by an
isotropic average of $\bm{l}_{i}^{-1}$,
\begin{equation}
  l_{0,\alpha\beta}^{-1} = \frac{1}{3} \left( 2l_{\perp}^{-1} +
    l_{\parallel}^{-1}\right) \,\delta_{\alpha\beta}.
\end{equation}
Thus we obtain
\begin{eqnarray}
  F_{\text{el}} &=& \frac{\mu}{2}\sum_{i=1}^{N}\left[ \left( \lambda
      ^{2}+2\lambda^{-1}\right) -\delta\left( \lambda^{2}-\lambda^{-1}\right)
    \left( \frac{3}{2}(\bm{m}\cdot\bm{n}_{i})^{2}-\frac{1}
      {2}\right) \right] \nonumber\\ 
  &=& \frac{\mu N}{2}\left( \lambda^{2}+\frac{2}{\lambda}\right) -\frac
  {\mu\delta}{3}\left( \lambda^{2}-\frac{1}{\lambda}\right) 
  \sum_{i=1}^{N} \sum_{\mu,\nu}M_{\mu\nu}S_{i}^{\mu\nu},
  \label{fel}%
\end{eqnarray}
where we have introduced the tensor%
\begin{equation}
  M_{\mu\nu}=\frac{1}{2}(3m^{\mu}m^{\nu}-\delta^{\mu\nu}),
\end{equation}
and the parameter%
\begin{equation}
  \delta=\frac{2l_{\perp}^{-1} - 2l_{\parallel}^{-1}}{2l_{\perp}^{-1}
    + l_{\parallel}^{-1}}, \qquad 0 \le \delta \le 1,
\end{equation}
with $\delta=0$ in the isotropic case, and
$\delta=1$ in the limit of largest anisotropy. The first term on the
right-hand side of equation\ (\ref{fel}) is the classical rubber free energy,
\begin{eqnarray}
  f_{\text{rub}} = \frac{\mu}{2} \left(\lambda^2 +\frac{2}{\lambda}\right).
\end{eqnarray}
Finally, the partition function may be written
\begin{equation}
  Z=\sum_{\{\bm{n}_{i}\}}\exp\left( -\beta E_{\text{eff}}\right) ,
\end{equation}
with the effective energy
\begin{eqnarray}
  E_{\text{eff}}=-\frac{A}{N}\sum_{1\leq i<j\leq
    n}\sum_{\mu,\nu=x,y,z}   S_{i}^{\mu\nu}S_{j}^{\mu\nu} + 
  \frac{4}{9} B \sum_{i=1}^{n}\sum_{\mu,\nu=x,y,z}M_{\mu\nu}S_{i}
  ^{\mu\nu}+Nf_{\text{rub}}, 
\label{effec-en}
\end{eqnarray}
where we have defined
\begin{eqnarray}
  B = \frac{3}{4} \, \mu \,\delta \left(\lambda^2 -\frac{1}{\lambda}
  \right). 
\end{eqnarray}
Given the entropic origin of the elastic contribution to
the free energy, the linear shear modulus should depend linearly on
temperature,
\begin{eqnarray}
  \mu=n_{\text{s}}k_{\text{B}}T,
\end{eqnarray}
where $n_{\text{s}}$ is the number of strands in the polymer chain per unit volume \cite{warner03}.

We further assume that the microscopic nematic directors are subjected
to random fields, which originate from the distribution of local
anisotropy axes generated at the time of cross-linking. These
interactions may be represented by the energy term
\cite{fridrikh99,golubovic89,fridrikh97,yu98}
\begin{equation}
  E_{\text{rf}}=-\frac{\gamma}{2}\sum_{i=1}^{N}\sum_{\mu,\nu=x,y,z}H_{i}%
  ^{\mu\nu}S_{i}^{\mu\nu},
\end{equation}
with%
\begin{equation}
  H_{i}^{\mu\nu}=\frac{1}{2}\left( 3h_{i}^{\mu}h_{i}^{\nu}-\delta^{\mu\nu
    }\right) ,
\end{equation}
where $\gamma$ is an energy parameter, and $h_{i}^{\mu}$ is the $\mu$-th
component of the unit vector $\bm{h}_{i}$. We assume that $\left\{
  \bm{h}_{i}\right\} $ is a set of independent and identically
distributed quenched random variables with probability distribution
\begin{eqnarray}
  P(\bm{h})=\left\{
    \begin{array}{ll}%
      c/2, & \text{ for }\bm{h}=(0,0,\pm1),\\
      (1-c)/4, & \text{ for }\bm{h}=(\pm1,0,0),\,(0,\pm1,0),
    \end{array}
  \right. \label{discreted}%
\end{eqnarray}
with $0 \le c \le 1$. Let us choose the symmetry axis of the mesogen
units along the $z$ direction.  Then the parameter $c$ is related to
the degree of anisotropy of the mesogens attached to the network
cross-links, so that $c=1/3$ for samples formed in the isotropic state,
and $c>1/3$ for samples formed in the nematic state. Note that we have assumed the shape tensor $\bm{l}_{0,i}^{-1}$ be fixed as an
isotropic average of $\bm{l}_{i}^{-1}$, and chosen to include the network-heterogeneity history in the random-field interaction only.

\section{Calculation of the free energy}

Effects of a fixed external stress $\sigma$ will be taken into account
by means of the partition function
\begin{equation}
  Y(\{\bm{h}_i\}) = \int_0^\infty d\lambda \, e^{\beta \sigma \lambda}
  \sum_{\{\bm{n}_i\}} \exp \left[-\beta
    \left(E_{\text{eff}}+E_{\text{rf}}\right)\right], 
\end{equation}
for a given a configuration of random fields. The free energy density
is given by%
\begin{equation}
  f=-\frac{1}{\beta} \lim_{N\rightarrow \infty} \frac{1}{N} \ln Y 
  = f_{\text{rub}}-\sigma\lambda-\frac{1}{\beta}\lim_{N\rightarrow\infty} 
  \frac{1}{N}\ln\sum_{\{\bm{n}_{i}\}}\exp\left( -\beta E\right) ,
\end{equation}
where $f$ should be a minimum with respect to $\lambda$ and
\begin{equation}
  E=-\frac{A}{2N}\sum_{\mu,\nu}\left(
    \sum_{i=1}^{N}S_{i}^{\mu\nu}\right)^{2}-\frac{4}{9}B \sum_{i=1}^{N}
  \sum_{\mu,\nu}M_{\mu\nu}S_{i}^{\mu\nu } -
  \frac{\gamma}{2}\sum_{\mu,\nu}\sum_{i=1}^{N}H_{i}^{\mu\nu}S_{i}^{\mu\nu},
\end{equation}
where we have discarded terms of order smaller than $N$.

Using a set of standard Gaussian integral transformations
\cite{salinas93} we decouple the interactions between different
particles,
\begin{eqnarray}
  \sum_{\{\bm{n}_i\}}\exp\left(-\beta E\right)&=&\int\left[dQ \right]
  \exp \left(-\frac{N\beta A}{2} \sum_{\mu,\nu} Q_{\mu\nu}^{2} \right)
  \\ \nonumber &&\times \prod_i \left\{ \sum_{\{\bm{n}\}} \exp\left[
      \beta \sum_{\mu,\nu}S_{i}^{\mu\nu} \left( AQ_{\mu\nu} +
        \frac{4}{9}BM_{\mu\nu} + \frac{\gamma}{2}H_{i}^{\mu\nu}\right)
    \right] \right\}, 
\end{eqnarray}
where $\left[ dQ\right] =\prod_{\mu\nu}\sqrt{\beta AN/2\pi}%
dQ_{\mu\nu}$. Performing the sum over the
orientations $\bm{n}$ of a single particle we obtain
\begin{eqnarray}
  f &=& f_{\text{rub}}+\frac{B}{3}-\sigma\lambda-\frac
  {1}{\beta}\ln2 - \beta^{-1}\lim_{N\rightarrow\infty}\frac{1}{N}
  \ln\int\left[
    dQ\right] \exp\left[ -\frac{N\beta A}{2}\left( \sum_{\mu,\nu}Q_{\mu\nu}%
      ^{2}+\Tr \mathbb{Q}\right) \right]
  \nonumber \\ && \times
  \exp\left\{ \sum_{i=1}^{N}\left[ -\frac{\beta\gamma}{4}\Tr %
      \mathbb{H}_{i}+\ln\left( \sum_{\mu} e_\mu (\bm{h}_i)\right) \right]
  \right\},
\end{eqnarray}
where
\begin{eqnarray}
  e_\mu (\bm{h}_i) =\exp \left[\beta \left(\frac{3A}{2}Q_{\mu\mu}%
      +\frac{3 \gamma}{4}H_{i}^{\mu\mu}+ Bm_{\mu}^{2} \right) \right].
  \label{emu}
\end{eqnarray}
Invoking the law of large numbers, we have
\begin{equation}
  \lim_{N\rightarrow\infty}\frac{1}{N}\sum_{i=1}^N\left\{
    -\frac{\beta\gamma} {4}\Tr \mathbb{H}_{i}+\ln\left(
      \sum_{\mu} e_\mu (\bm{h}_i) \right) \right\} 
  =-\frac{\beta\gamma}{4}\left\langle
    \Tr \mathbb{H}\right\rangle _{h}+\left\langle \ln\left(
      \sum_{\mu} e_\mu (\bm{h}) \right) \right\rangle_{h},
\end{equation}
where $\left\langle ...\right\rangle _{h}$ denotes the expectation
value with respect to the random-field variables, from which we see
that the free energy is self-averaging. Carrying out the integration
using Laplace's method we arrive at
\begin{equation}
  f= f_{\text{rub}}+\frac{B}{3}- \sigma\lambda-\frac
  {1}{\beta}\ln2 - \beta^{-1} \max L(Q_{\mu\nu}),
\end{equation}
where $Q_{\mu\nu}$ maximizes the functional
\begin{equation}
  L=-\frac{\beta A}{2}\left( \Tr \mathbb{Q}^{2}+\Tr %
    \mathbb{Q}\right) +\left\langle \ln\left( \sum_{\mu}e_\mu
    \right) \right\rangle _{h}.\label{ldef}%
\end{equation}

The condition for $L$ to be stationary with respect to $Q_{\mu\nu}$
leads to the equations of state for the order parameters
\begin{equation}
  Q_{\mu\nu}=\frac{1}{2} \left( 3\left<\frac{e_\mu}{\sum_\alpha e_\alpha}
    \right>_h - 1\right) \delta_{\mu\nu}.
  \label{qmunu}
\end{equation}
Notice that $\Tr \mathbb{Q}=0$. The condition for the free energy to
be stationary with respect to $\lambda$ leads to the equation of state
for the distortion,
\begin{equation}
  \lambda=\frac{1}{\lambda^{2}}+\frac{\sigma}{\mu}+\frac{\delta}{2}\left(
    2\lambda+\frac{1}{\lambda^{2}}\right) \sum_{\mu}m_{\mu}^{2}Q_{\mu\mu},
  \label{lminf}%
\end{equation}
where we have used the result \eqref{qmunu}. Using these equations of
state we may rewrite the free-energy density as
\begin{equation}
  f=f_{\text{rub}}+\frac{B}{3}-\sigma\lambda-\frac{1}{\beta}
  \ln2+\frac{A}{2}\Tr \mathbb{Q}^{2} -\frac{1}{\beta}\left\langle
    \ln\left( \sum_{\mu} e_\mu \right) \right\rangle
  _{h}. \label{fe}
\end{equation}

To make a closer contact with experiments on liquid crystals, we use
the standard diagonal parametric form of the traceless matrix
$\mathbb{Q}$ appropriate for the nematic ordering along the $z$
direction,
\begin{equation}
  \mathbb{Q}=\left(
    \begin{array}
      [c]{ccc}%
      -\displaystyle\frac{S+\eta}{2} & 0 & 0\\
      0 & -\displaystyle\frac{S-\eta}{2} & 0\\
      0 & 0 & S
    \end{array}
  \right) . \label{qseta}%
\end{equation}
The nematic order parameters $S$ and $\eta$ characterize the isotropic
phase ($S=\eta=0$), the uniaxial phase ($S \ne 0, \eta=0$) and the
biaxial phase ($S \ne 0, \eta \ne 0$). From the equation of state
\eqref{qmunu} we find
\begin{eqnarray}
  S=\frac{3}{2} \left<\frac{e_z}{\sum_\mu e_\mu}\right>_h
  -\frac{1}{2},
  \label{seqnorep}
\end{eqnarray}
and%
\begin{eqnarray}
  \eta=\frac{3}{2}\left\langle \frac{e_y-e_x}{\sum_\mu e_\mu}
  \right\rangle _{h}. 
\end{eqnarray}

The simplest version of the MSZ model defined by Eqs.
(\ref{emseq1}-\ref{nzwanz}) lacks the proper symmetry to describe a
stable nematic biaxial phase. To check if this behavior is robust
after including elasticity and random fields, we have considered the
case where the strain direction is perpendicular to the
\textit{chosen} axis of symmetry of the order parameter. In fact,
there has been a lot of debate in the literature about the soft (or
semisoft) response of NEs being related to the onset of biaxial
behavior \cite{warner03,verwey96,warner02,ye09}. While the description
of such phenomena is beyond the scope of this paper, we emphasize that
a numerical inspection of the equations of state have led us to
conclude that biaxial symmetry remains absent for a large range of
parameters. Hence we shall take $\eta=0$ and consider the strain
direction parallel to the $z$ axis, $\bm{m} = (0,0,1)$. This is a
reasonable assumption since the coupling between elastic and
orientational degrees of freedom provides an easy axis of symmetry for
this system. Thus the free-energy density \eqref{fe} becomes
\begin{equation}
  f=f_{\text{rub}}+\frac{1}{3}B-\sigma\lambda-\frac{1}{\beta}
  \ln2+\frac{3A}{4}S^2 -\frac{1}{\beta}\left\langle
    \ln\left(\sum_\mu e_\mu\right) \right\rangle _{h},
  \label{feqnorep}
\end{equation}
where
\begin{eqnarray}
  e_{\mu} (\bm{h})&=&\exp \left\{\beta \left[-\frac{3}{4} A S
      +\frac{3}{8} \gamma \left(3 h_{\mu}^2 - 1\right) \right] \right\},
  \quad \mu=x,y, \\
  e_z (\bm{h})&=&\exp \left\{\beta \left[\frac{3}{2} A S
      +\frac{3}{8} \gamma \left(3 h_z^2 - 1\right) + B \right] \right\},
\end{eqnarray}
and the equation of state for distortion \eqref{lminf} takes the form
\begin{equation}
  \lambda=\frac{1}{\lambda^{2}}+\frac{\sigma}{\mu}+\frac{\delta}{2}\left(
    2\lambda+\frac{1}{\lambda^{2}}\right) S.
  \label{leqnorep}%
\end{equation}

The results obtained thus far are valid for arbitrary random-field
distributions. Henceforth we limit ourselves to the discrete
distribution (\ref{discreted}). In this case the free-energy density
\eqref{feqnorep} is given by
\begin{eqnarray}
  f &=&f_{\text{rub}} +
  \frac{1}{3}B-\sigma\lambda-\frac{1}{\beta} \ln2 +
  \frac{3}{4}AS(S + 1) \\ \nonumber && -\frac{1}{\beta}\left\{
    c\ln\left[ 2 + e^{\beta(9 AS/4 + 9\gamma/8 + B)}\right]
    + (1-c)\ln\left[ 1 + e^{9\beta\gamma/8} + e^{\beta( 9 AS/4
        + B)}\right]   \right\},   
   \label{fedisc}
\end{eqnarray}
and the equation of state for the order parameter \eqref{seqnorep} becomes
\begin{equation}
  S =\frac{3}{2}\left[ \frac{c\, e^{ \beta (9 AS/4+9
        \gamma/8+B)} }{2+e^{\beta (9 AS/4+9 \gamma/8+ B)} }
    + \frac{(1-c)e^{ \beta(9 AS/4+ B)} }{1+e^{9\beta\gamma/8}
      +e^{\beta(9 AS/4+B)} } \right] -\frac{1}{2}.
\label{sdisc}
\end{equation}

\section{Thermodynamic results}

Without loss of generality, we assume $n_{\text{s}}=1$ and
$\delta=0.5$, since other choices lead to qualitatively similar
results. Let us write down the equations in terms of which we performed
numerical calculations. We express the energy in units of $A$ and the
temperature in units of $k_B/A$. The free-energy density
\eqref{fedisc} is given explicitly as
\begin{eqnarray}
  f&=&\frac{3}{4}S(S + 1) + T \Bigg\{
    \frac{1} {8}\left(5 \lambda^{2}-7\lambda^{-1}\right) -\ln 2
     -  c\ln\left[ 2 + e^{9(2S + \gamma)/8T
          + 3(\lambda^2-\lambda^{-1})/8} \right] 
     \nonumber \\
     &&- (1-c)\ln\left[ 1 + e^{9\gamma/8T } + e^{9S/4T +
          3(\lambda^2-\lambda^{-1})/8} \right] \Bigg\}. 
 \label{flast}%
\end{eqnarray}
The equation of state for the order parameter \eqref{sdisc} takes the
form 
\begin{eqnarray}
 S&=&\frac{3}{2}\left[ \frac{ce^{9(2S + \gamma)/8T +
 3(\lambda^2-\lambda^{-1})/8}}{2 + e^{9(2S + \gamma)/8T +
 3(\lambda^2-\lambda^{-1})/8}} + \frac{(1-c)e^{9S/4T +
 3(\lambda^2-\lambda^{-1})/8}}{1 + e^{9\gamma/8T} + e^{9S/4T +
 3(\lambda^2-\lambda^{-1})/8}} \right] -\frac{1}{2},
 \label{slast}%
\end{eqnarray}
and the equation of state for the distortion \eqref{leqnorep} gives
\begin{eqnarray}
 \lambda=\frac{1}{\lambda^{2}} + \frac{\sigma}{T} + \frac{1}{4}\left(
 2\lambda + \frac{1}{\lambda^{2}}\right) S. 
 \label{llast}%
\end{eqnarray}

In the absence of applied stress ($\sigma=0$), the equation of state
\eqref{llast} can easily be solved for the distortion with the result
\begin{equation}
  \lambda=\left( \frac{1 + S/4}{1-S/2}\right) ^{\frac{1}{3}}.
\end{equation}
The distortion increases monotonically with $S$, for $0<S<1$. In
addition, $S=0$ implies $\lambda=1$, indicating that the strain is
driven by the orientational ordering only. For an arbitrary applied
stress ($\sigma \ge 0$) the equation of state \eqref{llast} is a cubic
equation in $\lambda$ with only one real and positive root given by
the formula
\begin{equation}
  \lambda = \frac{2 \sigma}{3(2-S)T} \left\{ 1 + 2\cosh
    \left[\frac{1}{3} \cosh^{-1} \left( 1 +
        \frac{27(2-S)^2(4+S)T^3}{32 \sigma^3} \right) \right]
  \right\},
\label{lambda}
\end{equation}
which is a monotonically increasing function of $S$.

In our numerical calculations we solved  Eq.  \eqref{slast} for the
order parameter $S$ using for $\lambda$ the result given by Eq.
\eqref{lambda}.

\subsection{Results  in the absence of disorder ($\gamma=0$)}

\begin{figure}[!ht]
\vspace{0.5cm}
 \begin{center}
 \includegraphics[width=10cm]{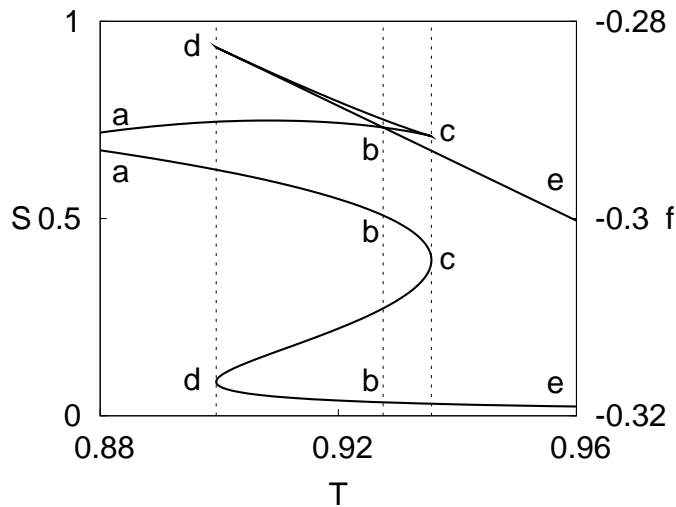}
 \end{center}
 \caption{Order parameter $S$ and free-energy density $f$ (upper
   curve) as a function of temperature for $\gamma=0$ and applied
   stress $\sigma=0.02$}
 \label{sfxt-g0}%
\end{figure}

In Fig. \ref{sfxt-g0} we plot the order parameter $S$ and the
free-energy density $f$ (upper curve) as a function of temperature for
applied stress $\sigma=0.02$ in the temperature range where a
first-order transition takes place.  The low temperature stable
solution $abc$ with larger order parameter $S$ will be called nematic
solution, whereas the high temperature solution $dbe$ with smaller
order parameter will be referred to as isotropic solution, even though
$S$ no longer vanishes as in the case of zero applied stress. The
branch $cd$ corresponds to an unstable solution. In the narrow
temperature interval $T_d < T < T_c$ both the nematic and the
isotropic solutions are stable, and we have to choose the one with
smaller free energy. The first-order transition between nematic and
isotropic phases occurs at the temperature $T_{\text{NI}}=T_b$ where
the free energies become equal.  We observe that, in agreement with
classical elasticity theory, the free energy changes linearly with
temperature in the isotropic phase \cite{warner03}.

\begin{figure*}[!ht]
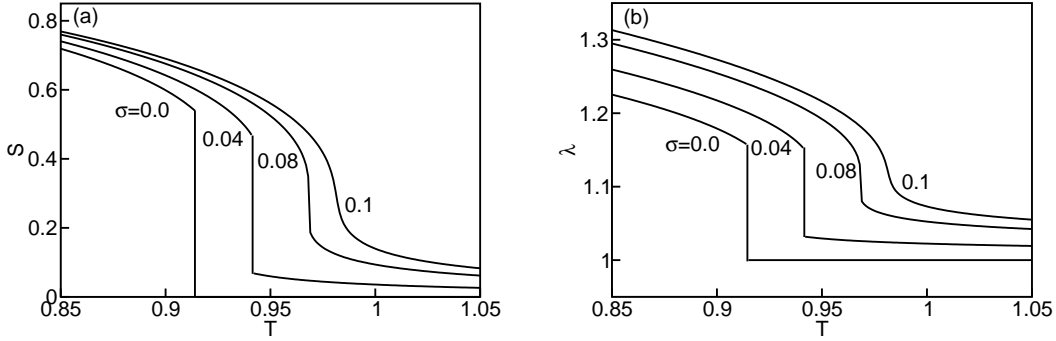

\begin{minipage}[b]{0.4\linewidth}
\includegraphics[width=\linewidth]{sxt_g0}
\end{minipage}
\hspace{0.5cm}
\begin{minipage}[b]{0.4\linewidth}
\includegraphics[width=\linewidth]{lxt_g0}
\end{minipage}
\caption{Nematic order parameter (a), and distortion factor (b) as a
  function of temperature, for $\gamma=0$ and various applied stresses
  $\sigma$.}%
\label{slxt-g0-sig}%
\end{figure*}

In Fig. \ref{slxt-g0-sig} we plot (a) the nematic order parameter $S$
and (b) the distortion factor $\lambda$ as a function of temperature
for several values of the applied stress $\sigma$.  As expected from
Eq. \eqref{lambda}, the graph of $\lambda$ follows closely that of
$S$. For small applied stress $\sigma$ the system undergoes a
first-order transition with a gap between the nematic and isotropic
solutions.  As the aligning stress $\sigma$ increases, the gap
decreases until the critical point is reached, beyond which there is
no phase transition. This behavior has been predicted by de Gennes in
the mid-seventies \cite{gennes75} before nematic elastomers were
proven to be chemically feasible.  Experimentally, however, no
first-order transition is observed down to the limit of zero applied
stress. The jump in the first-order transition is smoothed out, being
replaced by a continuous but quick variation of the order parameter.
This fact has been interpreted theoretically as being due to the
anisotropic distribution of random fields
\cite{petridis06,selinger04}.

\subsection{Effects of disorder ($\gamma>0$)}

\begin{figure*}[!ht]
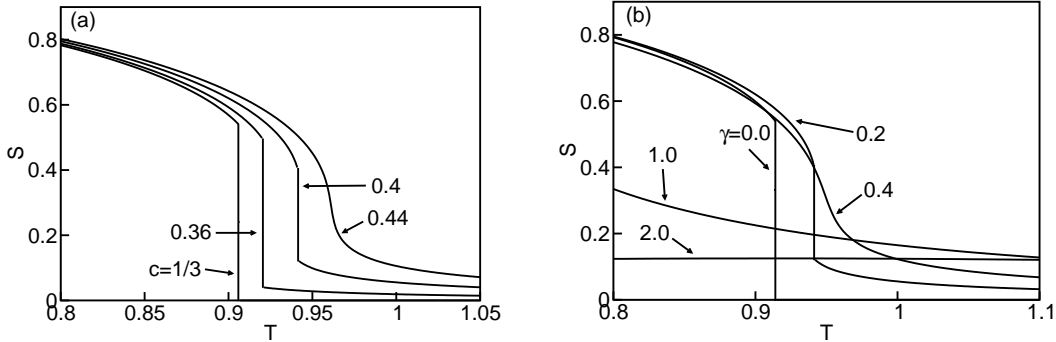

\vspace{0.5cm}
 \begin{minipage}[b]{0.4\linewidth}
 \includegraphics[width=\linewidth]{sxt_gam02_sig0}
 \end{minipage}
 \hspace{0.5cm}
 \begin{minipage}[b]{0.4\linewidth}
 \includegraphics[width=\linewidth]{sxt_sig0_c04_g00-01}
 \end{minipage}
 \caption{Nematic order parameter as a function of temperature for $\sigma=0$.
 (a) $\gamma=0.2$ and various values of  $c$. 
 (b) $c=0.4$ and various values of $\gamma$.}%
\label{slxt-g02-sig0}%
\end{figure*}

Let us examine how the random fields affect the nematic-isotropic
transition. In Fig.  \ref{slxt-g02-sig0}a we show the nematic order
parameter as a function of temperature for $\sigma=0$, $\gamma=0.2$
and several values of $c$.  We observe that the gap between the
nematic and isotropic solutions decreases as $c$ increases from
$c=1/3$ to $c = 0.422$, disappearing above this value of $c$.  In
Fig. \ref{slxt-g02-sig0}b the nematic order parameter is shown as a
function of temperature for $c=0.44$ and several values of $\gamma$.
Again, the gap between the nematic and isotropic solutions decreases
as $\gamma$ increases from $\gamma=0$ to $\gamma =0.263$, and the
transition disappears above this value of $\gamma$.

\begin{figure}[!ht]
\vspace{0.5cm}
 \begin{center}
 \includegraphics[width=10cm]{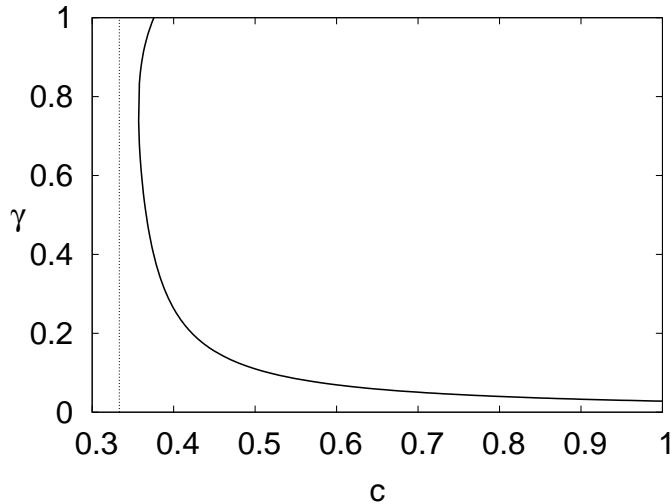}
 \end{center}
 \caption{The curve for $\sigma=0$ above and to the right of which
   there is no first-order transition. The vertical dotted line
   corresponds to $c=1/3$.}
 \label{gammaxc}%
\end{figure}

These results show that anisotropic ($c > 1/3$) distribution of the
random fields of sufficient strength ($\gamma>0$) is necessary to
smooth out the isotropic-nematic transition, in agreement with the
numerical simulations of Selinger and Ratna \cite{selinger04}.  Fig.
\ref{gammaxc} shows, for zero applied stress ($\sigma=0$), the curve in
the $\gamma$--$c$ plane above and to the right of which there is no
first-order transition.

\subsection{Results for  an isotropic disorder ($c=1/3$, $\gamma > 0$ )}

\begin{figure}[!ht]
\vspace{0.5cm}
 \begin{center}
 \includegraphics[width=8cm]{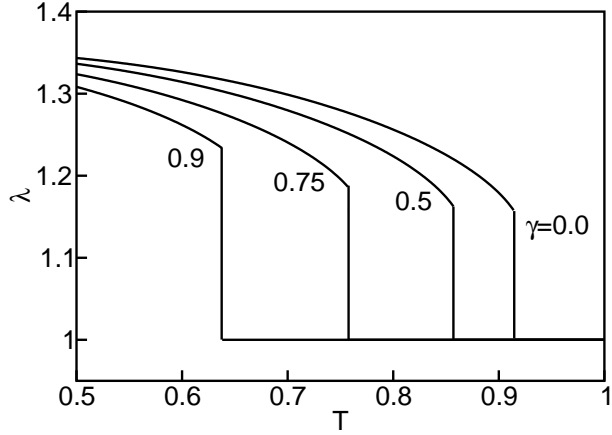}
 \end{center}
 \caption{Distortion factor $\lambda$ as a function of temperature for
   $\sigma=0$, $c=1/3$, and $\gamma$ between $0$ and~$1$.}%
 \label{lxt-c33-sig0-g}%
\end{figure}

We recall that an isotropic distribution of random fields $c=1/3$
represents samples formed in the isotropic state, with the random
stresses coming from the cross-linked network without preferred
direction. According to Fig.  \ref{gammaxc}, in this case a first-order
transition occurs in zero applied stress ($\sigma=0$) for any value of
the random-field strength $\gamma$.  To illustrate this fact, we plot
in Fig. \ref{lxt-c33-sig0-g} the distortion factor $\lambda$ as a
function of temperature in the absence of external stress ($\sigma=0$)
for several values of the disorder parameter $\gamma$.  The nematic
phase decreases with increasing $\gamma$, but the first-order
transition persists showing no evidence of the experimentally observed
smoothed out nematic-isotropic transition in samples formed in the
nematic state.

\begin{figure*}[!ht]
\vspace{0.5cm}
 \begin{minipage}[b]{7.5cm}
 \includegraphics[width=\linewidth]{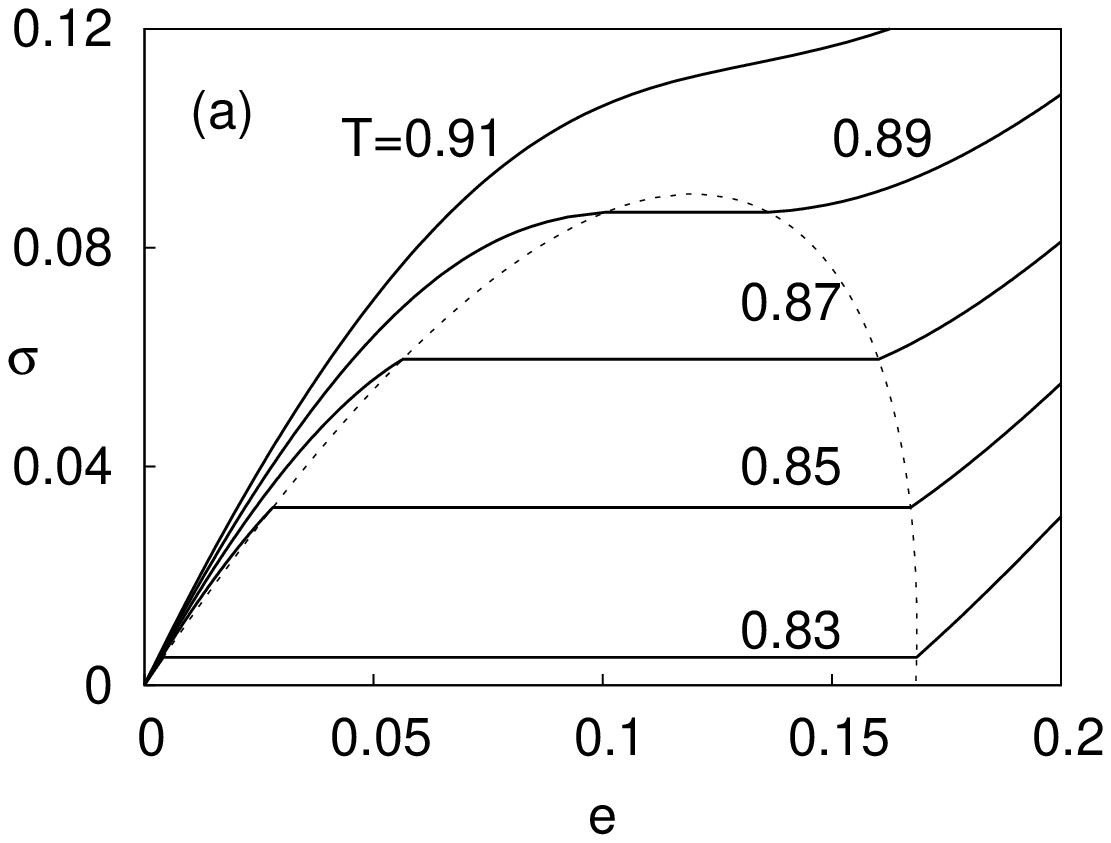}
 \end{minipage}
 \hspace{0.5cm}
 \begin{minipage}[b]{7.5cm}
 \includegraphics[width=\linewidth]{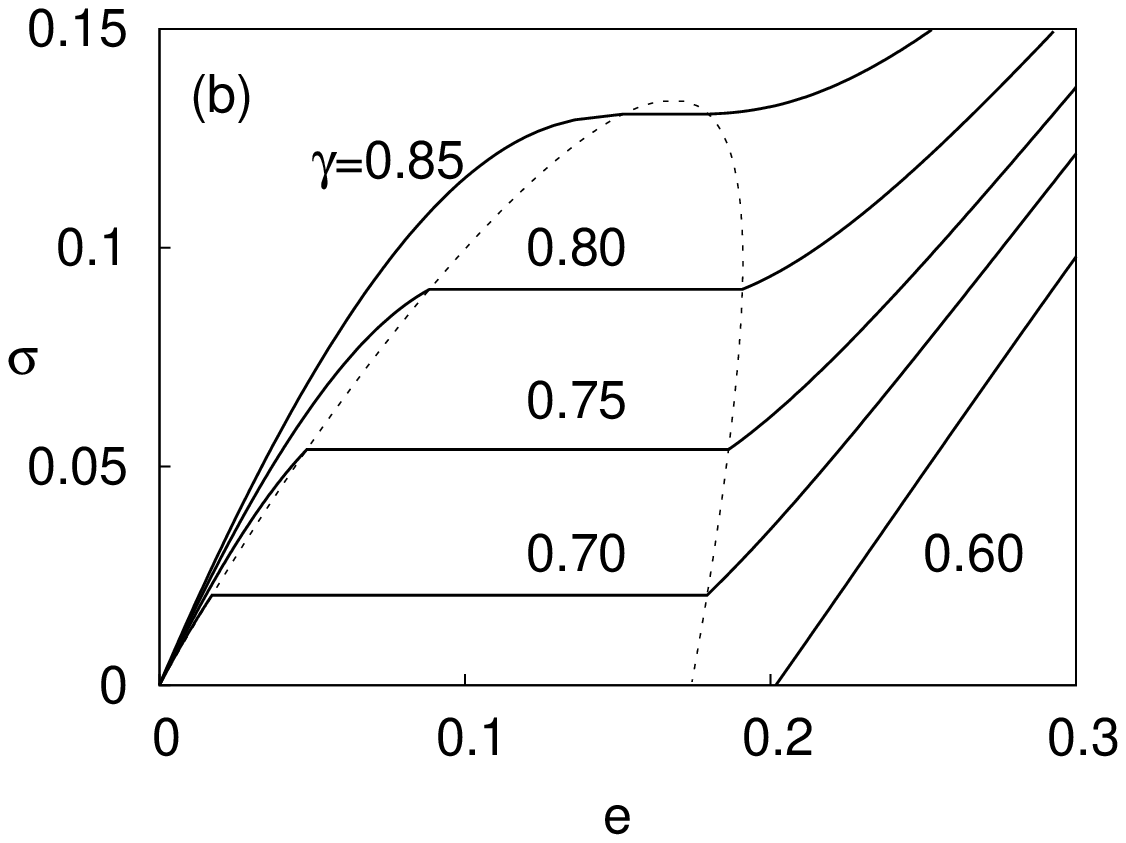}
 \end{minipage}
 \caption{Stress-strain curves for $c=1/3$. (a) Isotherms for
   $\gamma=0.6$. (b) Iso-$\gamma$ curves for $T=0.8$.}
\label{sigxe}%
\end{figure*}

In the presence of the applied stress ($\sigma > 0$), the first-order
transition is smoothed out by a sufficiently large temperature or
disorder strength. This can be seen in the stress-strain curves, where
the strain $e$ is related to the distortion by the equation $\lambda =
1+e$. In Fig. \ref{sigxe}a we plot isotherms for $\gamma=0.6$ and in
Fig. \ref{sigxe}b we plot iso-$\gamma$ curves for $T=0.8$. It is clear
from these figures that the temperature and the disorder strength have
similar effect on the system.  For sufficiently low temperature ($T <
T_c$) or disorder strength ($\gamma < \gamma_c$), the stress grows
monotonically with strain in the isotropic phase up to the first-order
transition to the nematic phase. At the transition the two phases
coexist and the strain is independent of stress.  The stress-strain
\textquotedblleft plateau\textquotedblright\ is then followed by a
stress growth at larger strains.  This behavior agrees with the
experimental scenario of a typical transition between polydomains and
a monodomain in NEs.  As the temperature or the disorder strength
increases, the coexistence curve shrinks until the critical point is
reached at $T=T_c$ or $\gamma = \gamma_c$.  The whole phase diagram
resembles a typical $P$-$V$ diagram of simple fluids.

\section{Discussions}

\begin{figure}[!ht]
\vspace{0.5cm}
 \begin{center}
 \includegraphics[width=8cm]{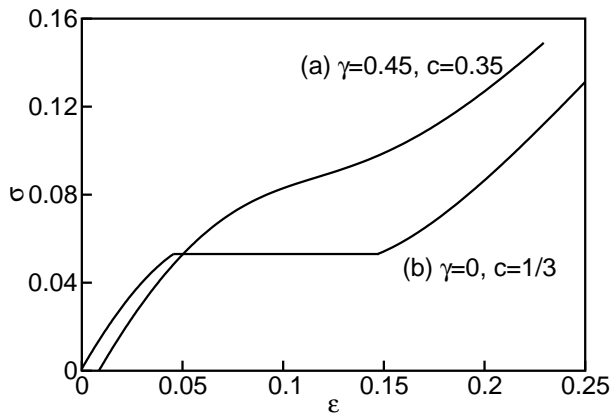}
 \end{center}
 \caption{Stress-strain curves for $T=0.95$. (a) $\gamma=0.45$ and $c=0.35$. (b) $\gamma=0$ and $c=1/3$.}
 \label{urayama-comparison}%
\end{figure}

Recent experiments have shown that NEs crosslinked in the isotropic state display a well-defined plateau at the stress-strain curve, for substantially lower critical stresses, in comparison with NEs crosslinked in the nematic state \cite{urayama09}. Optical microscopy observations suggest this behavior be attributed to larger memory effects for NEs formed in the nematic state. We may then expect that $\gamma$ is not independent, but should increase with $c$, according to the language of our model. Now it is not difficult to find appropriate values of $\gamma$ and $c$ satisfying this restriction, and in general agreement with the experimental results. We show in Figure \ref{urayama-comparison} two stress-strain curves for $T=0.95$. For curve (a), we consider samples crosslinked in the isotropic state ($c=1/3$), which implies no memory effect ($\gamma =0$). In this case, the stress-strain curve presents a well-defined plateau for low critical stress. For curve (b), we consider samples crosslinked in the nematic state ($c=0.35$), which should imply strong memory effects ($\gamma=0.45$). Note that the stress-strain curve characterizes a broad polydomain-monodomain transition for higher critical stress, in good agreement with the experimental results by Urayama et al. \cite{urayama09}.

In conclusion, we have introduced a simple mean-field lattice model to describe the
behavior of nematic elastomers. This model combines the
Maier-Saupe-Zwanzig theory of liquid crystals
\cite{maier58,henriques97,carmo10} and the lattice version, due to
Selinger and Ratna \cite{selinger04}, of the Warner-Terentjev theory
of elasticity \cite{warner03}. We performed detailed calculations for
a large range of parameters, with the inclusion of the effects of a
quenched distribution of random fields. A stress-strain coexistence
curve may be obtained for systems cooled below a freezing temperature,
which is analogous to the $P$-$V$ diagram of a simple fluid, with the
disorder strength playing the role of temperature. Below a critical
stress, the characteristic tie lines resemble the experimental
stress-strain plateau, and may be interpreted as signatures of a
polydomain-monodomain transition. In the monodomain case, we show that
random-field disorder may soften the first-order transition between
nematic and isotropic phases, provided the samples are formed in the
nematic state. Beyond general agreement with some previous findings,
we hope our results may motivate further experimental work on the
stress-strain coexistence curve of nematic elastomers.

\begin{acknowledgments}
 We acknowledge the financial support of the Brazilian agency CNPq.
\end{acknowledgments}


%

\end{document}